\documentclass[english,aps,prl,longbibliography,groupedaddress,superscriptaddress,reprint]{revtex4-1}
\pdfoutput=1
\usepackage[T1]{fontenc}
\usepackage[latin9]{inputenc}
\usepackage{textcomp}
\usepackage{amstext}
\usepackage{wasysym}
\usepackage{graphicx}
\usepackage{hyperref}
\usepackage[figure,figure*]{hypcap}
\usepackage{etoolbox}

\newcounter{myequation}
\makeatletter
\@addtoreset{equation}{myequation}
\makeatother

\newcounter{myfigure}
\makeatletter
\@addtoreset{figure}{myfigure}
\makeatother

\makeatletter
 
 \@ifundefined{textcolor}{}
 {%
   \definecolor{BLACK}{gray}{0}
   \definecolor{WHITE}{gray}{1}
   \definecolor{RED}{rgb}{1,0,0}
   \definecolor{GREEN}{rgb}{0,1,0}
   \definecolor{BLUE}{rgb}{0,0,1}
   \definecolor{CYAN}{cmyk}{1,0,0,0}
   \definecolor{MAGENTA}{cmyk}{0,1,0,0}
   \definecolor{YELLOW}{cmyk}{0,0,1,0}
 }

\usepackage{placeins}

\makeatother

\usepackage{babel}
\begin{document}

\title{A scanning transmon qubit for strong coupling circuit quantum electrodynamics}

\author{W. E. Shanks}

\email{wshanks@princeton.edu}

\selectlanguage{english}%

\author{D. L. Underwood}

\author{A. A. Houck}

\affiliation{Department of Electrical Engineering, Princeton University, Olden
Street, Princeton, NJ 08550}

\date{\today}

\maketitle

\section*{Main body}

Like a quantum computer designed for a particular class of problems,
a quantum simulator enables quantitative modeling of quantum systems
that is computationally intractable with a classical computer. Quantum
simulations of quantum many-body systems have been performed using
ultracold atoms\textsuperscript{\citealp{bloch2012quantum}} and trapped
ions\textsuperscript{\citealp{blatt2012quantum}} among other systems.
Superconducting circuits have recently been investigated as an alternative
system in which microwave photons confined to a lattice of coupled
resonators act as the particles under study with qubits coupled to
the resonators producing effective photon-photon interactions\textsuperscript{\citealp{houck2012onchipquantum}}.
Such a system promises insight into the nonequilibrium physics of
interacting bosons but new tools are needed to understand this complex
behavior. Here we demonstrate the operation of a scanning transmon
qubit and propose its use as a local probe of photon number within
a superconducting resonator lattice. We map the coupling strength
of the qubit to a resonator on a separate chip and show that the system
reaches the strong coupling regime\textsuperscript{\citealp{chiorescu2004coherent,wallraff2004strongcoupling}}
over a wide scanning area.

Over the past decade, the study of quantum physics using superconducting
circuits has seen rapid advances in sample design and measurement
techniques\textsuperscript{\citealp{schoelkopf2008wiringup,clarke2008superconducting,martinis2009superconducting}}.
A great strength of superconducting qubits compared to other promising
candidates is that they are fabricated using standard lithography
procedures which allow fine tuning of qubit properties and make scaling
up the fabrication to devices with many qubits straightforward. Circuit
quantum electrodynamics (CQED) is an active branch of quantum physics
research in which one or more qubits are strongly coupled to a superconducting
coplanar waveguide resonator (CPWR) which is used to control and readout
the state of the qubits\textsuperscript{\citealp{blais2004cavityquantum,wallraff2004strongcoupling}}.
A prerequisite for most interesting CQED applications is that the
system reach the strong coupling regime in which the rate at which
the qubit and the resonator exchange an excitation exceeds the excitation
decay rate. 

In addition to the CQED architecture's promise as a quantum computing
platform, recent theoretical work has focused on using a CQED lattice,
a network of coupled resonators each coupled to its own qubit, as
a non-equilibrium quantum simulator\textsuperscript{\citealp{houck2012onchipquantum}}.
One particularly interesting prediction for CQED lattice systems is
a cross-over from a superfluid-like state to an insulating state as,
for example, the coupling between the qubits and their resonators
is increased\textsuperscript{\citealp{greentree2006quantum,hartmann2006strongly,angelakis2007photonblockadeinduced}},
similar to the superfluid-Mott insulator quantum phase transition
which has been observed in ultracold atom systems\textsuperscript{\citealp{greiner2002quantum,bloch2012quantum}}.
More exotic phenomena including analogs of the fractional\textsuperscript{\citealp{hayward2012fractional}}
and anomalous\textsuperscript{\citealp{petrescu2012anomalous}} quantum
Hall effects and of Majorana physics\textsuperscript{\citealp{bardyn2012majoranalike}}
have also been considered. However, while preliminary steps have been
taken to build such a CQED lattice system\textsuperscript{\citealp{underwood2012lowdisorder}},
both establishing and probing its expected many-body states remain
major experimental challenges.

\begin{figure}[!b]
\includegraphics{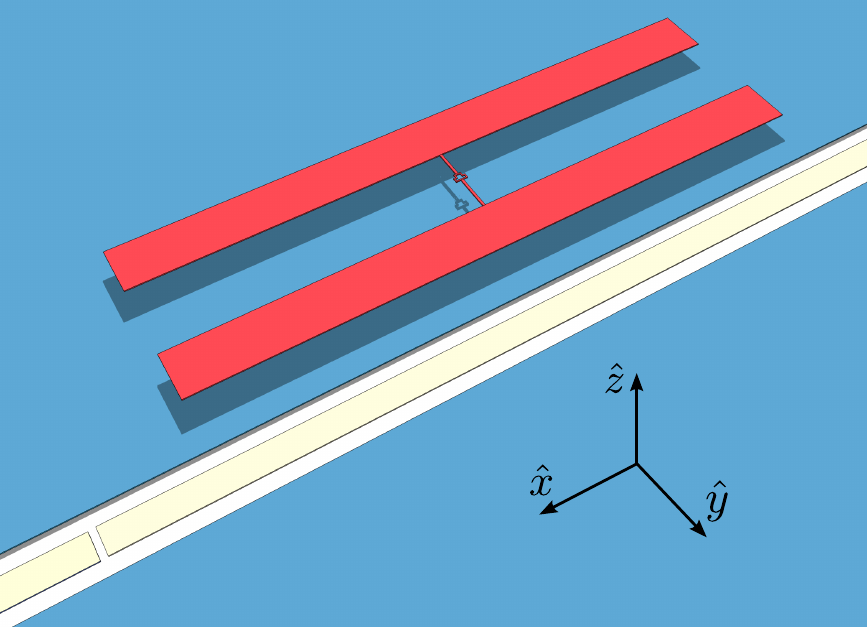}

\caption{\label{fig:sample_images}Scanning transmon qubit. The qubit (red)
is composed of two $40\times500\,\text{\textmu m}$ aluminum islands
separated by a $60\,\text{\textmu m}$ gap and connected by a thin
wire containing a Josephson tunnel barrier. In the qubit discussed
here, the tunnel barrier is formed by two aluminum oxide barriers
in the two arms of a loop in the middle of the wire. The qubit is
pictured $11\,\text{\textmu m}$ above a portion of a CPWR. The resonator's
center pin (yellow) is $21\,\text{\textmu m}$ wide and is separated
from the surrounding ground plane (blue) by $12\,\text{\textmu m}$
gaps. The ends of the resonator are formed by gaps in the center pin
like the one shown in the figure. The resonator has a length of $7,872\,\text{\textmu m}$
which corresponds to a half-wave resonance of $7.6\,\text{GHz}$ in
the absence of the qubit. The resonator center pin has no curves in
over half its length, so that the qubit remains aligned as it scans
over this part of the resonator. To make them more visible, the qubit
and resonator have been thickened, but all other dimensions are to
scale with the samples discussed here.}

\end{figure}

The simplest method of probing a microwave circuit is to measure transmission
between two of its ports. However, in the case of a CQED lattice such
a measurement gives only limited insight into the detailed behavior
of photons in the interior. Additional information could be obtained
by measuring transmission while locally perturbing the interior of
the sample, as has been done to image the coherent flow of electrons
in two-dimensional electron gas systems\textsuperscript{\citealp{eriksson1996cryogenic,topinka2000imaging}}.
This local perturbation requires a new scanning probe tool, such as
the one demonstrated here. Besides just perturbing the lattice, a
scanning probe qubit can be used to measure the photon number of individual
lattice sites following a protocol used to measure photons in non-scannable
cavities\textsuperscript{\citealp{johnson2010quantum}}. One benefit
of a scanning qubit in this case is its ability to measure the photon
number of interior lattice sites. Measurements of outer resonators,
those most easily accessed by a measurement circuit fabricated on
the same chip as the lattice, would be difficult to interpret due
to edge effects.

While our primary focus is on using the scanning qubit in the context
of CQED lattice-based quantum simulation, we note that scanning qubits
have been studied previously for other applications. Scanning nitrogen-vacancy
(NV) center qubits in diamond have been demonstrated to be sensitive
local probes of magnetic field\textsuperscript{\citealp{rondin2012nanoscale,maletinsky2012arobust}}.
A diamond NV center has also been coupled to a scanning photonic crystal
cavity\textsuperscript{\citealp{englund2010deterministic}}. While
this experiment did not reach the strong coupling regime, the scanning
cavity enhanced spontaneous emission from the NV center and allowed
its position to be determined with greater spatial resolution.

The scanning qubit described here (Fig. \ref{fig:sample_images})
is a transmon design consisting of two aluminum islands connected
by a thin aluminum wire interrupted by an aluminum oxide tunnel barrier\textsuperscript{\citealp{koch2007chargeinsensitive}}.
The tunnel barrier provides a large nonlinear inductance, which together
with the capacitance between the two islands, makes the transmon behave
as a nonlinear LC oscillator whose lowest two energy states can be
used as a qubit. The transmon design is well suited for scanning because
it couples to CPWRs capacitively and requires no physical connections.
The qubit chip was mounted face down to a cryogenic three-axis positioning
stage and positioned over a separate chip containing a niobium CPWR
with a half-wave resonance at $7.6\,\text{GHz}$. In order to avoid
direct contact between the resonator and the qubit, pads of photoresist
$7\,\text{\ensuremath{\mu}m}$ thick were deposited on the corners
of the qubit chip. The sample holder was mounted to a dilution refrigerator
which operated at temperatures $\apprle35\,\text{mK}$. 

The main result presented here is the measurement of the strength
$g$ of the coupling between the resonator and the qubit as a function
of qubit position. Following ref. \citealp{koch2007chargeinsensitive},
the Hamiltonian $\hat{H}$ describing the coupled resonator-qubit
system can be approximated by
\begin{equation}
\hat{H}=h\nu_{r}\left(\hat{a}^{\dagger}\hat{a}+\frac{1}{2}\right)+\frac{h\nu_{q}}{2}\hat{\sigma}_{z}+\frac{hg}{2}\left(\hat{a}\hat{\sigma}^{+}+\hat{a}^{\dagger}\hat{\sigma}^{-}\right)\label{eq:hamiltonian}
\end{equation}
with $\nu_{r}$ and $\nu_{q}$ the resonator and qubit frequencies
respectively. In this expression, $\hat{a},\hat{a}^{\dagger}$ are
the creation and annihilation operators associated with photons in
the resonator and $\hat{\sigma}^{+}$, $\hat{\sigma}^{-}$, and $\hat{\sigma}_{z}$
are the Pauli spin matrices associated with the qubit when treated
as a two-level system. On resonance $(\nu_{q}=\nu_{r})$, the first
two excited states of the system are $(|0\uparrow\rangle\pm|1\downarrow\rangle)/\sqrt{2}$
with corresponding energies $h\nu_{r}\pm hg$ above that of the ground
state $|0\downarrow\rangle$ where $|nq\rangle$ is the state with
$n$ photons in the resonator and the qubit in state $q$ with $\downarrow$
($\uparrow$) representing the qubit ground (excited) state. When
driven with a microwave excitation, transitions to each of these excited
states are allowed, resulting in two peaks in the low power transmission
spectrum. These peaks are separated in frequency by $2g$, called
the vacuum Rabi splitting.

The frequency $\nu_{r}$ of the resonator depends on its capacitance
to its ground plane which is decreased when the qubit chip is brought
into close proximity. In order to ensure that resonance was possible
at every qubit position, the qubit was fabricated with a pair of tunnel
barriers integrated into a loop in place of a single tunnel barrier.
By varying the flux through this loop with a magnet coil incorporated
into the positioner, the qubit frequency $\nu_{q}$ could be varied
from a maximum value of $12.1\,\text{GHz}$ to close to zero\textsuperscript{\citealp{koch2007chargeinsensitive}}.
We note that, although here the flux loop's only purpose is to make
the qubit energy tunable, such a loop can also be operated as a sensitive
local magnetometer in a scanning SQUID microscope\textsuperscript{\citealp{huber2008gradiometric}}.

\begin{figure*}
\includegraphics{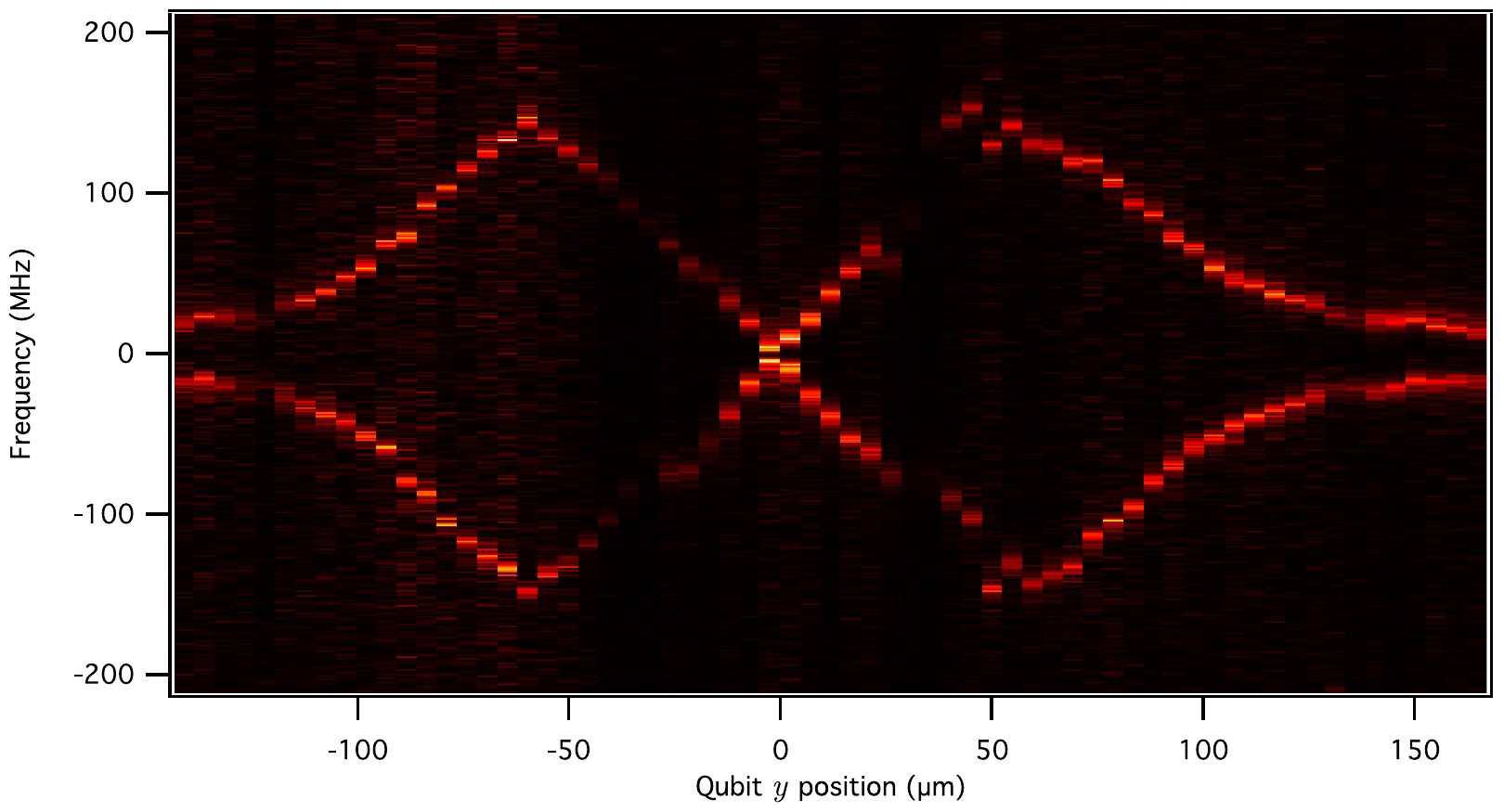}

\caption{\label{fig:resonant_transmission_vs_y}Resonant transmission spectra.
Transmission spectra taken with the qubit in resonance with the resonator
are shown for a series of qubit positions. The vertical axis represents
the difference between the frequency of the applied microwave drive
and the resonator frequency. The shifted peak locations can also be
interpreted as values of the coupling strength $g$ which is equal
to the peak separation. We take the $y$ origin to be the point of
smallest peak separation which we interpret as the position where
the qubit is centered over the resonator. The suppressed transmission
at positions $y=\pm30$ and $\pm125\,\text{\textmu m}$ is due to
coupling between the resonator and a parasitic resonance in the metal
frame of the qubit chip (see Supplemental Discussion).}
\end{figure*}

Fig. \ref{fig:resonant_transmission_vs_y} shows the transmission
spectra of the resonator for a sequence of regularly spaced qubit
positions along the $y$ axis perpendicular to the resonator. At each
position, the current through the magnet coil was adjusted to bring
the qubit into resonance, where the single transmission peak of the
resonator was transformed into two peaks of equal height, clearly
demonstrating strong coupling between the scanning qubit and the resonator.
The position scan shows two regions of large peak separation symmetric
about a position with nearly no peak separation which we set as the
origin. In coupling to the resonator the transmon behaves as a dipole
antenna. Because the two islands of the qubit are identical, by symmetry
no coupling is expected when the qubit is centered above the resonator
at $y=0$. The points of maximum peak separation occur at $y\approx\pm50\,\text{\textmu m}$
where one of the two islands is centered over the resonator. At these
points, the observed coupling strength $g\approx140\,\text{MHz}$
was well into the strong coupling regime $g>\kappa,T_{1}^{-1}$ where
the qubit relaxation time $T_{1}=2.6\,\text{\textmu s}$ was determined
by time-domain measurements (see Supplemental Methods) and the photon
escape rate $\kappa=13\,\text{MHz}$ was determined from the resonator
linewidth. The photon escape rate was chosen to be large in order
to increase the rate of data acquisition.

\begin{figure*}
\includegraphics{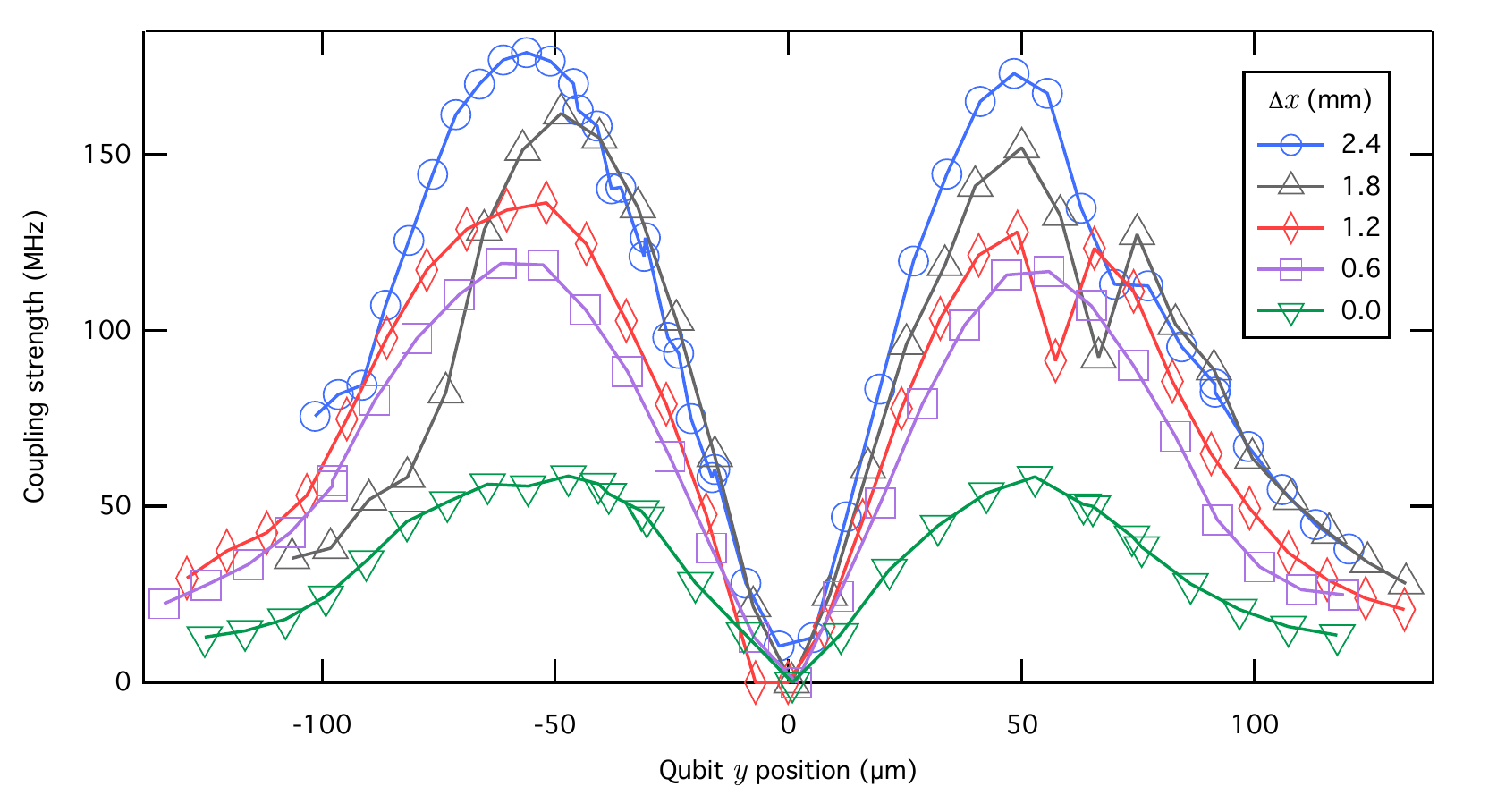}

\caption{\label{fig:g_vs_xy}Coupling strength $g$ versus qubit position.
Traces of $g$ versus qubit $y$ position are shown for five qubit
$x$ positions spaced $600\,\text{\textmu m}$ apart from each other.
The traces correspond to scans like the one shown in Fig. \ref{fig:resonant_transmission_vs_y}.
Each value of $g$ was determined by fitting several transmission
spectra taken at values of magnetic flux for which the qubit frequency
was close to that of the resonator. The sign of $\Delta x$ was chosen
so that with increasing $\Delta x$ the qubit moved from the electric
field node at the center of the resonator towards the electric field
antinode at its end.}
\end{figure*}

Scans of resonant transmission versus $y$ position like Fig. \ref{fig:resonant_transmission_vs_y}
were repeated at five positions along the length of the resonator
(the $\hat{x}$ direction) with a spacing of $600\,\text{\textmu m}$.
The coupling strengths $g$ extracted from fits to the transmission
spectrum at each qubit position are plotted in Fig. \ref{fig:g_vs_xy}.
The coupling strength increases as the qubit moves from the voltage
node at the center of the resonator to the antinode at its end but
exhibits the same shape for its $y$ dependence at each $x$ position.

\begin{figure}
\includegraphics{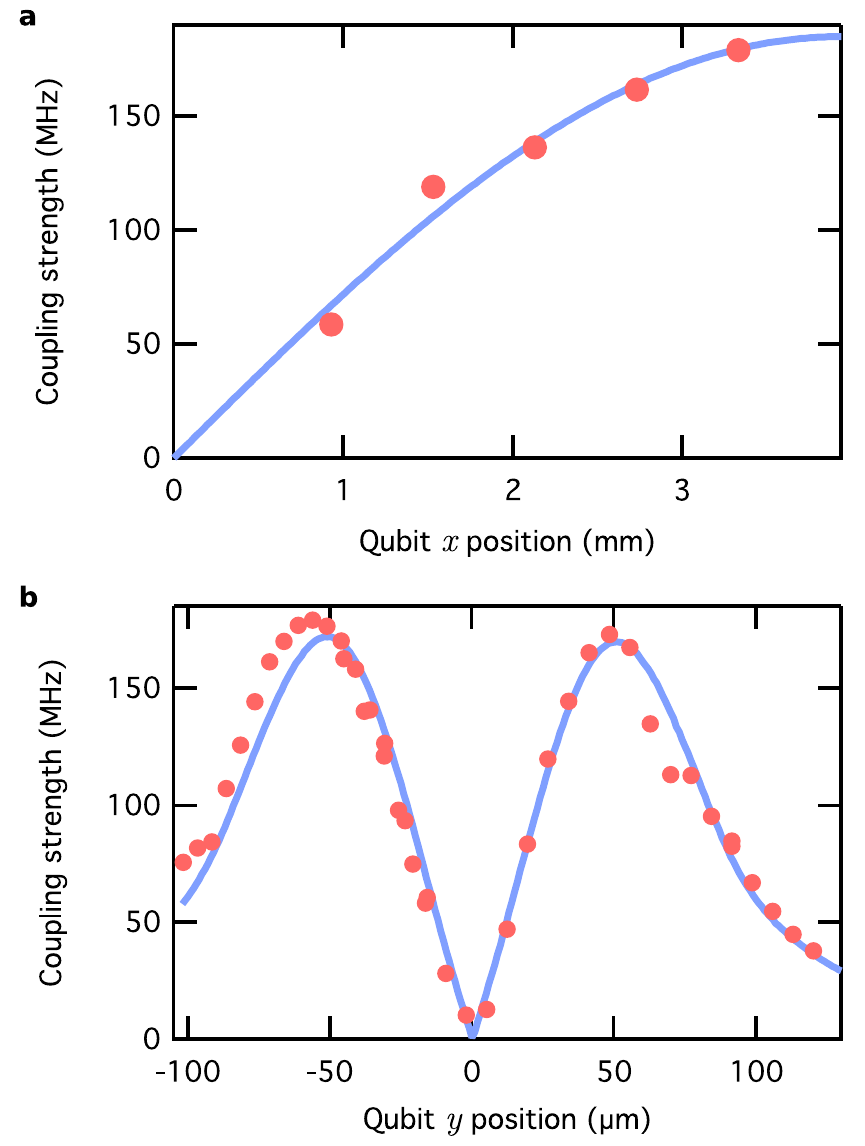}

\caption{\label{fig:g_fits}Quantitative analysis of measured coupling strength.
\textsf{\textbf{a,}} The maximum coupling strength $g$ of each trace
in Fig. \ref{fig:g_vs_xy} is plotted versus $x$ position along with
a fit to the expected sinusoidal dependence. The $x$ origin represents
the midpoint of the resonator. The offset of the data points from
the resonator midpoint was determined by the fit (see Supplemental
Methods). The other fit parameter, the maximum coupling strength the
end of the resonator, was found to be $185\,\text{MHz}$. \textsf{\textbf{b,}}
The largest trace of $g$ vs. $y$ in Fig. \ref{fig:g_vs_xy} is replotted
along with a fit to the form expected from finite element modeling
of the qubit-resonator system's capacitance matrix. The fitting function
uses the resonator frequency $\nu_{r}$, the system geometry, and
the qubit $x$ position determined in panel \textsf{\textbf{a}} as
fixed inputs and treats the qubit height $z$, found to be $11\,\text{\textmu m}$,
as its only free parameter.}
\end{figure}

We now consider the data in Fig. \ref{fig:g_vs_xy} more quantitatively.
The coupling strength $g$ is proportional to the electric field at
the position of the qubit when a single photon is present in the resonator.
Panel \textbf{a} of Fig. \ref{fig:g_fits} shows the maximum value
of $g$ for each $x$ position along with a fit to the expected sinusoidal
dependence of the electric field strength along the $x$ direction.
The fit provides an absolute reference for the relative $x$ positions
quoted in Fig. \ref{fig:g_vs_xy}. With the $x$ position known, we
can compare the measurement data with the $g(x,y,z)$ obtained from
finite element simulations (see Supplemental Methods). Panel \textbf{b}
of Fig. \ref{fig:g_fits} shows the values of $g$ observed for $x=3,330\,\text{\textmu m}$
from the resonator midpoint along with the simulation results which
show good agreement. The simulation height $z=11\,\text{\textmu m}$
is somewhat larger than the $7\,\text{\textmu m}$ thickness of the
photoresist pads on the corners of the qubit chip but corresponds
to a misalignment between the qubit and resonator chips of $\sim0.1^{\circ}$
over the $2\,\text{mm}$ from the edge of the qubit chip to the qubit's
location at the chip center.

In conclusion, we have observed strong coupling between a scanning
transmon qubit and a CPWR. Because strong coupling can be reached,
anything possible with stationary qubits now becomes possible with
a moveable probe opening the door for a wide array of applications
in the field of CQED. Such a scanning qubit makes possible quantum
measurements of superconducting circuits with spatial resolution.
In addition to the scanning measurements of a lattice CQED system
discussed earlier, we note that the system studied here demonstrates
\emph{in situ} tuning of the coupling strength $g$ which is often
a desirable capability experimentally. While the scanning qubit's
coupling can not be tuned on the timescale of the coherence time like
some previously demonstrated circuit designs\textsuperscript{\citealp{bialczak2011fasttunable,srinivasan2011tunable}},
it does not require flux or current biasing of the system. For example,
for the lattice CQED system mentioned above, a lattice of resonators
on one chip could be coupled to a lattice of qubits on a second chip
allowing the coupling between each resonator-qubit pair to be tuned
together as one chip is scanned over the other. An array of qubits
could also be used to measure the statistics of qubit coherence by
scanning the qubits one by one across a single measurement resonator.

\section*{Methods}

The qubit was fabricated using electron beam lithography and double-angle
shadow evaporation with controlled oxidation of $30$ and $100\,\text{nm}$
layers of aluminum onto a $4\times4\,\text{mm}$ sapphire chip. The
$0.5\times1.0\,\text{mm}$ crashpads on the corners of the chip were
made with photolithography of SU-8 2005 photoresist. The resonator
was defined by photolithography and acid etch ($\text{H}_{2}\text{O}$,
HF, and $\text{HNO}_{3}$ in a 7.5:4:1 ratio) of a $200\,\text{nm}$
film of niobium on a $14\times14\,\text{mm}$ sapphire chip. 

The qubit chip was glued with methyl methacrylate to the tip of a
highly conductive copper rod mounted to the cryogenic positioning
stage (Attocube ANPx340/RES, ANPz101/RES). The resonator chip was
mounted to a copper patterned circuit board with silver paste and
aluminum wire bonds which connected the input and output transmission
lines to coaxial lines. Wire bonds were only placed around the edge
of the chip outside the footprint of the qubit chip. The wiring scheme
of the coaxial lines was the same as that described in ref. \citealp{dicarlo2009demonstration}.

All values of qubit position were determined by potentiometric measurements
of resistive position encoders integrated into the positioning stage.
Individual position readings had an uncertainty of $0.4\,\text{\textmu m}$,
and overall the position readings drifted by $1.8\,\text{\textmu m}$
per $100\,\text{\textmu m}$. A typical movement of the positioning
stage heated the refrigerator from its base temperature of $\apprle15\,\text{mK}$
to over $85\,\text{mK}$. In order to reduce measurement time, most
measurements were taken with the refrigerator in the range between
$25$ and $35\,\text{mK}$ which took only a couple minutes to reach
after moving the stage.

\section*{Acknowledgements}

This work was supported by DARPA under grant \#N66001-10-1-4023. DLU
is supported by a fellowship from the NSF (DGE-1148900).

\section*{Author Contributions}

A. A. H. conceived and supervised the experiment. D. L. U. designed
and built the sample holder and fabricated the samples. W. E. S. designed
and fabricated the samples, performed the measurements and data analysis,
and wrote the manuscript. All authors discussed the results and implications
and commented on the manuscript at all stages.

\section*{Competing Financial Interests}

The authors acknowledge no competing financial interests.

\begin{quotation}
\newpage{}
\end{quotation}

\FloatBarrier

\renewcommand{\thefigure}{S\arabic{figure}}
\renewcommand{\theequation}{S\arabic{equation}}

\stepcounter{myequation}
\stepcounter{myfigure}

\section*{Supplementary Methods}

\subsection*{Fitting functions}

Non-linear least-squares fitting routines were used to determine the
coupling strength $g$ from transmission data and to produce the curves
shown in Fig. \ref{fig:g_fits}. Here we briefly describe the functions
used in each case.

\subsubsection*{Transmission measurements}

\begin{figure*}
\includegraphics{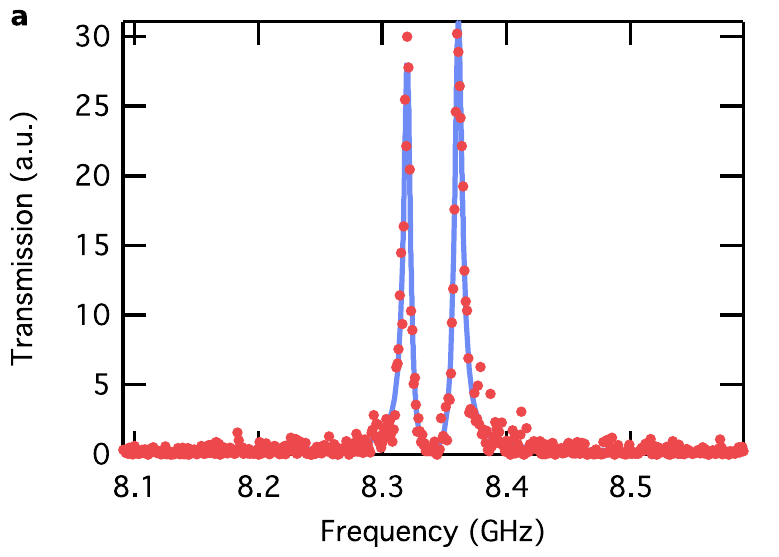}\includegraphics{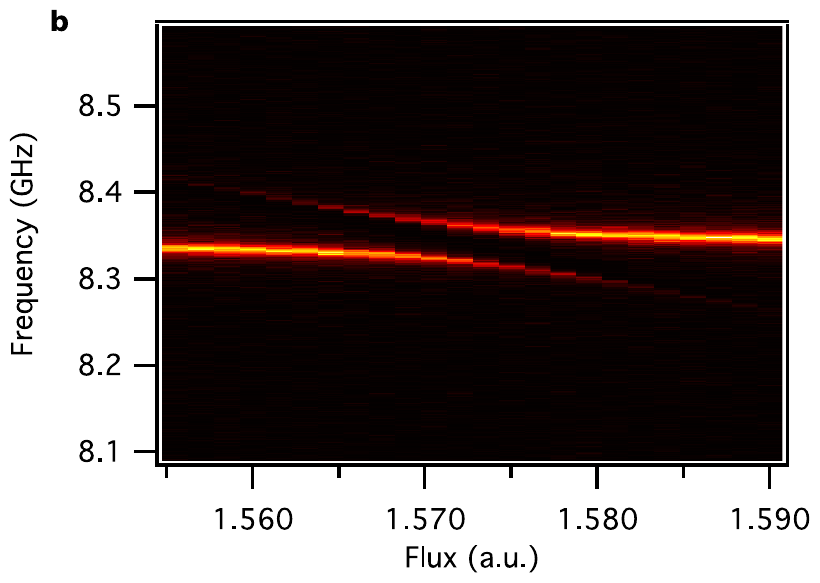}

\caption{\label{fig:transmission_fit}Fitting resonator transmission. \textsf{\textbf{a,}}
The transmission at $y=146\,\text{\textmu m}$ in Fig. \ref{fig:resonant_transmission_vs_y}
is plotted in arbitrary units along with a fit to equation (\ref{eq:transmission}).
The fit coefficients are $A=57$, $B=0.2$, $g=20\,\text{MHz}$, $\kappa=14\,\text{MHz}$,
$\nu_{r}=8.342\,\text{GHz}$, and $\nu_{q}=8.339\,\text{GHz}$. \textsf{\textbf{b,}}
Transmission spectra recorded at the same position as panel \textsf{\textbf{a}}
are plotted versus flux in arbitrary units as the qubit passes through
resonance. Averaging the coefficients from fits to the transmission
at each flux value gives $A=59\pm3$, $B=0.30\pm0.05$, $g=20.4\pm0.3\,\text{MHz}$,
$\kappa=12.8\pm0.8\,\text{MHz}$, and $\nu_{r}=8.342\pm0.001\,\text{GHz}$
where the errors represent the standard deviation of the coefficients.}
\end{figure*}

Transmission measurements were made in the low power limit for which
the rate of photons entering the resonator was less than the escape
rate, so that the resonator occupancy was less than one photon on
average. In this case, the transmission spectrum contains peaks at
frequencies $\nu_{\pm}$ corresponding to transitions from the ground
state $|0\downarrow\rangle$ to the states $|+\rangle$ and $|-\rangle$
which possess a $|1\downarrow\rangle$ component and are located at
energies $h\nu_{+}$ and $h\nu_{-}$ above the ground state. The states
and their frequencies are found by diagonalizing the Hamiltonian given
in equation (\ref{eq:hamiltonian}):
\begin{widetext}
\[
|\pm\rangle=\frac{\frac{\Delta}{2}\pm\sqrt{\left(\frac{\Delta}{2}\right)^{2}+g^{2}}}{\sqrt{g^{2}+\left(\frac{\Delta}{2}\pm\sqrt{\left(\frac{\Delta}{2}\right)^{2}+g^{2}}\right)^{2}}}|0\uparrow\rangle+\frac{g}{\sqrt{g^{2}+\left(\frac{\Delta}{2}\pm\sqrt{\left(\frac{\Delta}{2}\right)^{2}+g^{2}}\right)^{2}}}|1\downarrow\rangle
\]
\end{widetext}
and

\begin{equation}
\nu_{\pm}=\frac{\nu_{r}+\nu_{q}\pm\sqrt{4g{}^{2}+\Delta^{2}}}{2}\label{eq:mode_frequencies}
\end{equation}
with $\Delta=\nu_{q}-\nu_{r}$. The peak amplitudes are proportional
to the probabilities $w_{\pm}=|\langle1\downarrow|\pm\rangle|^{2}$
of a photon being measured in the states $|\pm\rangle$. The peak
linewidths $\gamma_{\pm}$ are equal to the decay rates of the qubit
($T_{1}^{-1}$) and the photon ($\kappa$) weighted by the probability
of measuring a qubit excitation and a photon respectively: $\gamma_{\pm}=w_{\pm}\kappa+(1-w_{\pm})T_{1}^{-1}$.
We take transmission peaks as following a lorentzian lineshape and
use the following form to fit the resonator transmission:
\begin{equation}
S_{21}\left(\nu\right)=B+A\left|\sum_{\pm}w_{\pm}l\left(\nu,\nu_{\pm},\gamma_{\pm}\right)\right|^{2}\label{eq:transmission}
\end{equation}
where $A$ is the overall amplitude accounting for all attenuation
and amplification in the measurement circuit, $B$ is the background
of the detector, and $l(\nu,\nu_{0},\gamma)$ is the complex lorentzian
centered at $\nu_{0}$ with width $\gamma$:
\[
l\left(\nu,\nu_{0},\gamma\right)=\left(1-i\frac{\nu-\nu_{0}}{\gamma/2}\right)^{-1}.
\]
When fitting for $g$, the parameters $A$, $B$, $\kappa$, $\nu_{q}$,
$\nu_{r}$, and $g$ were allowed to vary, while $T_{1}$ was held
fixed to the value obtained from coherence measurements. Fig. \ref{fig:transmission_fit}
shows the result of fitting one of the transmission spectra from Fig.
\ref{fig:resonant_transmission_vs_y}. Also shown is a plot of transmission
versus flux as the qubit passes through resonance. For the coupling
strength values shown in Fig. \ref{fig:g_vs_xy}, similar flux scans
were taken and the plotted values of coupling strength were obtained
by averaging the coupling strength values obtained from fitting the
transmission at each flux value.

\subsubsection*{Coupling strength versus $x$}

The voltage profiles of the modes of a CPWR with open boundary conditions
are sinusoidal along the length of the resonator with antinodes at
its ends. The coupling strength is proportional to the resonator voltage
with one photon present and so should follow this sinusoid. The maximum
coupling strength at each $x$ position shown in panel \textsf{\textbf{a}}
of Fig. \ref{fig:g_fits} was fit to the sinusoidal form:
\begin{equation}
g\left(\Delta x\right)=g_{\max}\sin\left(\pi\frac{\Delta x+x_{0}}{l_{r}}\right)\label{eq:gofx}
\end{equation}
where $l_{r}=7,872\,\text{\textmu m}$ is the resonator length and
$g_{\max}$ and $x_{0}$ were the fitting parameters. Here $\Delta x$
is the set of displacements in $x$ from the first $x$ position (i.e.
the values are $0\,\text{\textmu m}$; $600\,\text{\textmu m}$; $1,200\,\text{\textmu m}$;
etc.). In panel \textsf{\textbf{a}} of Fig. \ref{fig:g_fits}, the
measured coupling strengths and the fit are plotted versus $x=\Delta x+x_{0}$.

\subsubsection*{Coupling strength versus $y$}

In order to perform the fit of coupling strength versus $y$ shown
in panel \textsf{\textbf{b}} of Fig. \ref{fig:g_fits}, the following
expression for the coupling strength given in ref. \citealp{koch2007chargeinsensitive}
was used:
\[
g\left(x,y,z\right)=2\sqrt{\frac{2Z_{c}}{h}}m\left(x\right)\nu_{r}\beta\left(y,z\right)n_{01}\left(y,z,\nu_{r}\right)
\]
with the characteristic line impedance $Z_{c}$ taken to be $50\,\Omega$
and $m(x)$ the sinusoidal mode shape factor given in equation (\ref{eq:gofx}).
In order to describe the voltage division factor $\beta$ and the
transmon matrix element $n_{01}$, we first define $C_{jk}$ to be
the capacitance between components $j$ and $k$ and label the components
of the system with $a$ and $b$ for the two islands of the transmon,
$p$ for the resonator center pin, and $g$ for all other pieces of
metal (the two ground planes and the metal frame on the qubit chip).
The voltage division factor $\beta$ gives the fraction of the voltage
drop from the resonator center pin to ground that falls across the
two islands of the qubit. It can be written in terms of capacitance
coefficients as
\[
\beta=\frac{\left|C_{ap}C_{bg}-C_{bp}C_{ag}\right|}{C_{ab}\left(C_{a\Sigma}+C_{b\Sigma}\right)+C_{a\Sigma}C_{b\Sigma}}
\]
where $C_{x\Sigma}=C_{xp}+C_{xg}$. We find the matrix element $n_{01}$
by numerically diagonalizing the transmon Hamiltonian given in ref.
\citealp{koch2007chargeinsensitive}
\[
\hat{H}=4E_{C}\hat{n}^{2}-E_{J}\cos\hat{\varphi}
\]
to finds it eigenstates and eigenenergies and then evaluating $n_{01}=\langle0|\hat{n}|1\rangle$,
where $|0\rangle$ and $|1\rangle$ are the eigenstates with the two
lowest energies, $E_{0}$ and $E_{1}$. The charging energy $E_{C}=e^{2}/2C_{\Sigma}$
used in the calculation was calculated using the total capacitance
given by
\[
C_{\Sigma}=C_{ab}+\left(\frac{1}{C_{a}}+\frac{1}{C_{b}}\right)^{-1}.
\]
The qubit frequency $\nu_{q}$ is given by $(E_{1}-E_{0})/h$ and
is thus a function of $E_{C}$ and $E_{J}$. In calculating $n_{01}$,
$\nu_{q}(E_{C},E_{J})$ was numerically inverted to solve for $E_{J}(E_{C},\nu_{q})$
with $\nu_{q}$ set equal to $\nu_{r}$ since measurements of the
coupling strength were made with the qubit close to the resonator's
frequency.

In order to produce the fit shown in panel \textsf{\textbf{b}} of
Fig. \ref{fig:g_fits}, the coupling strength $g(x,y,z)$ was calculated
using the known values of $Z_{c}$ and $\nu_{r}$, the value of $x$
obtained from the fit in panel \textsf{\textbf{a}} of Fig. \ref{fig:g_fits},
and the values of the capacitances $C_{jk}$ found by finite element
analysis for a grid of $y$ and $z$ values with $1\,\text{\textmu m}$
spacing. The measured coupling strength versus $y$ was fit to the
$g(y,z)$ found by interpolating between the $y$ and $z$ grid points
with $z$ as the only free parameter. The finite element simulation
was then repeated with the fitted value of $z$ in order to produce
the curve shown in panel \textsf{\textbf{b}} of Fig. \ref{fig:g_fits}.
We note that at the fitted value of $z=11.0\,\text{\textmu m}$ the
charging energy $E_{C}=388\,\text{MHz}$ is similar to values used
in other CQED experiments and corresponds to a ratio of $E_{J}/E_{C}=59$,
within the transmon regime where the offset charge across the transmon
islands (not included in the Hamiltonian given above) may be ignored. 

By the use of alignment marks on the resonator and qubit chips, it
was possible to confirm that the misalignment between the two chips
was $3^{\circ}\pm1^{\circ}$. A misalignment of $3^{\circ}$ was used
in the finite element calculations for the capacitance coefficients.
Using a misalignment of $2^{\circ}$ ($4^{\circ}$) instead gave a
the fitted height $z$ of $11.1\,\text{\textmu m}$ ($10.7\,\text{\textmu m}$).

\subsection*{Coherence measurements}

Qubit coherence times ($T_{1}=2.6\pm0.3\,\text{\textmu s}$, $T_{2}^{*}=1.0\pm0.2\,\text{\textmu s}$)
were obtained using the techniques described in ref. \citealp{schreier2008suppressing}.
The measurements were made during the same cooldown and at the same
$x$ position as the data shown in Fig. \ref{fig:resonant_transmission_vs_y}.
For technical reasons, the measurements were made immediately after
the refrigerator was warmed up to $20\,\text{K}$ and then cooled
back down to its base temperature. The coherence measurements were
performed at $y=-113\,\text{\textmu m}$ ($g=31\,\text{MHz}$) in
Fig. \ref{fig:resonant_transmission_vs_y} with the qubit frequency
detuned $700\,\text{MHz}$ below the resonator.

\section*{Supplementary Discussion}

Here we provide some additional measurements and analysis related
to scanning the qubit over the resonator. In Fig. \ref{fig:resonant_transmission_vs_y},
the transmission is suppressed near $y=\pm30$ and $\pm125\,\text{\textmu m}$
due to the resonator's coupling to spurious modes. Because the behavior
of these modes is symmetric in qubit position, we believe them to
be caused by resonances between the resonator chip and a layer of
metal on the qubit chip that was patterned symmetrically. The extra
metal on the qubit chip was deposited for technical reasons and is
not needed for the functioning of the qubit. By redesigning the qubit
chip or resonator chip, these modes could be eliminated.

\begin{figure*}
\includegraphics{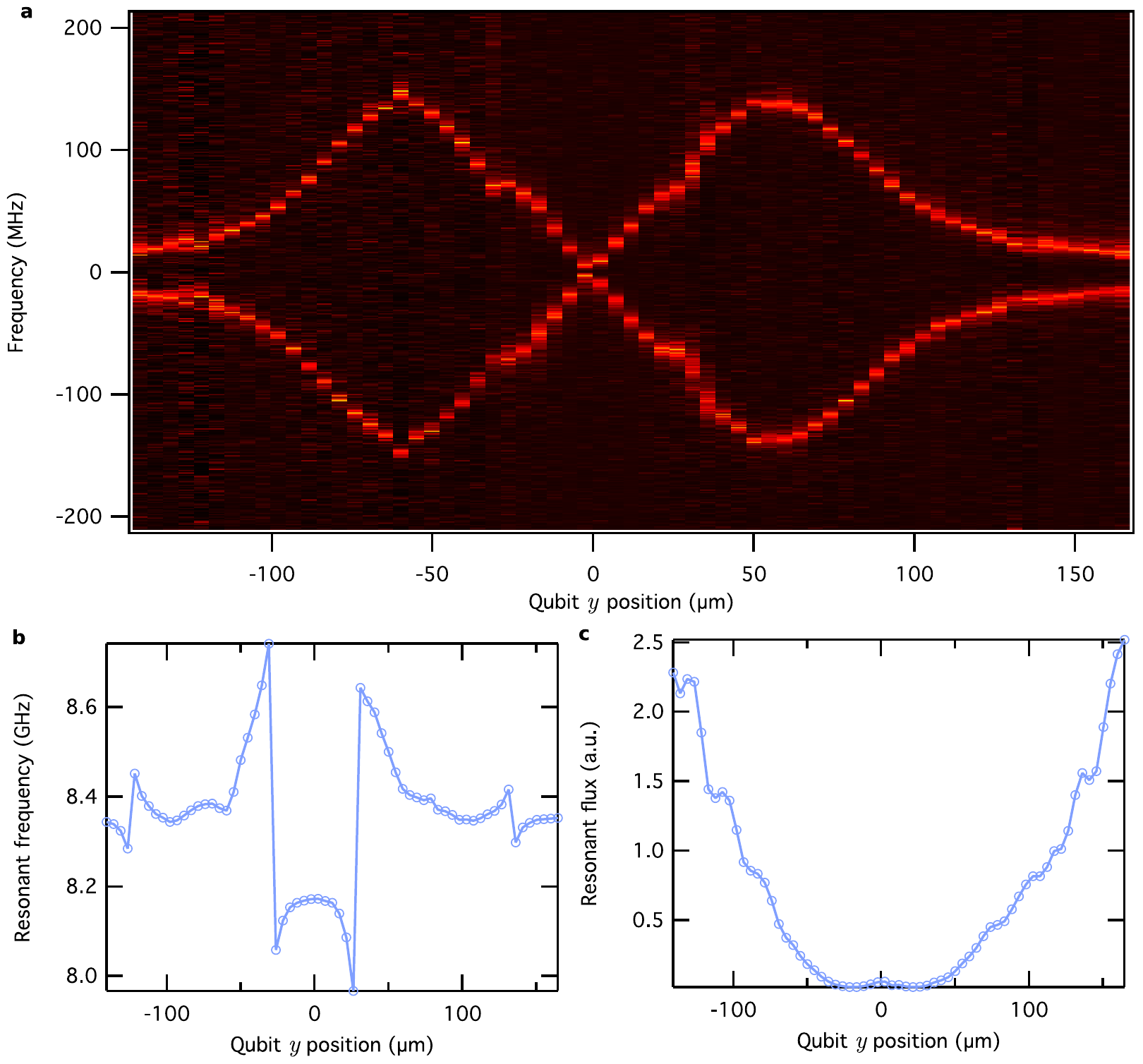}

\caption{\label{fig:scaled_resonant_transmission}Additional information about
Fig. \ref{fig:resonant_transmission_vs_y}. \textsf{\textbf{a,}} The
transmission spectra with the qubit tuned into resonance with the
resonator is plotted for a series of qubit positions. At each position,
the background is removed and the transmission is scaled so that the
maximum transmission is unity. The frequency of each spectrum is offset
so that the two peaks are centered around zero. \textsf{\textbf{b,}}
The resonator frequency is plotted for each qubit position of Fig.
\ref{fig:resonant_transmission_vs_y}. The frequency axis of each
spectrum in Fig. \ref{fig:resonant_transmission_vs_y} was offset
by the frequency plotted here. \textsf{\textbf{c,}} The total flux
(in arbitrary units) sourced by the magnet coil to bring the qubit
into resonance is plotted versus qubit position for the scan shown
in Fig. \ref{fig:resonant_transmission_vs_y}.}
\end{figure*}

In panel \textsf{\textbf{a}} of Fig. \ref{fig:scaled_resonant_transmission},
the data from Fig. \ref{fig:resonant_transmission_vs_y} is replotted
after removing the background and normalizing the transmission peaks
to unity in order to make the peaks near $y=\pm30$ and $\pm125\,\text{\textmu m}$
more visible. In Fig. \ref{fig:resonant_transmission_vs_y}, the origin
of the frequency axis was set to the location of the high power transmission
peak and the qubit frequency was tuned to produce two peaks of equal
height in transmission. For some positions near where the resonator
coupled to the spurious modes, these conditions produced two peaks
not centered around zero. In panel \textsf{\textbf{a}} of Fig. \ref{fig:scaled_resonant_transmission},
the frequency axis at each position has been shifted to center the
peaks around zero, so that the transmission peaks at neighboring positions
can be more easily compared.

The spurious modes appeared as lower and broader peaks in transmission
at frequencies that varied with position and did not vary with magnetic
flux. These modes were always present but only affected the measurement
of the resonator-qubit system when their frequencies were close to
the resonator frequency. When the frequency of one of these modes
was close to the resonator frequency, the coupling between the spurious
mode and the resonator resulted in two modes with excitations partially
of the resonator and partially of the spurious mode. The narrower
peak of the resulting two peaks in transmission was chosen to be the
resonator peak for the purpose of coupling to the qubit. Panel \textsf{\textbf{b}}
of Fig. \ref{fig:scaled_resonant_transmission} shows the frequency
of the chosen peak for each position in Fig. \ref{fig:resonant_transmission_vs_y}.
Jumps in the resonator frequency due to avoided crossings with spurious
modes are visible at the $y$ positions with low transmission in Fig.
\ref{fig:resonant_transmission_vs_y}.

The scan shown in Fig. \ref{fig:resonant_transmission_vs_y} was taken
on a separate cooldown from the scans shown in Fig. \ref{fig:g_vs_xy}.
The same resonator and qubit samples were used for both sets of measurements,
but the sample stage was disassembled in between the cooldowns. During
the cooldown in which data in Fig. \ref{fig:resonant_transmission_vs_y}
was taken, the qubit's $x$ position was not varied, so the absolute
$x$ position of the data is not known. However, using the maximum
value of $g$ from Fig. \ref{fig:resonant_transmission_vs_y} and
the curve shown in panel \textsf{\textbf{a}} of Fig. \ref{fig:g_fits}
to calibrate the $x$ position, one finds the data in Fig. \ref{fig:resonant_transmission_vs_y}
was taken at $x=2,116\,\text{\textmu m}$.

Changing the position of the qubit also affected the threading of
flux through the qubit's SQUID loop. Once the SQUID loop was moved
away from the gaps in the coplanar waveguide and positioned above
the superconducting ground plane, the amount of flux produced by the
magnet coil required to tune qubit into resonance increased rapidly
because the Meissner effect screened the magnetic field away from
the superconducting ground plane. In panel \textsf{\textbf{c}} of
Fig. \ref{fig:scaled_resonant_transmission}, the coil magnetic flux
that brought the qubit into resonance with the resonator is plotted
for each position of Fig. \ref{fig:resonant_transmission_vs_y}. The
impact of the magnetic field screening could be greatly reduced by
fabricating holes in the resonator's ground plane.

The strong dependence of the resonant magnetic flux on the qubit position
as well as the steepness of the slope of qubit frequency versus magnetic
flux (maximum qubit frequency $\nu_{q,\max}\sim12.1\,\text{GHz}$)
necessitated careful flux scanning at each qubit position in order
to locate resonance. Searching for resonance by monitoring the transmission
spectrum for an avoided crossing feature like the one shown in panel\textsf{\textbf{
b}} of Fig. \ref{fig:transmission_fit} would have required a long
measurement time at each qubit position. Instead, only the low power
transmission at the frequency $\nu_{r}$ of the high power transmission
peak was monitored as the flux was swept. 

\begin{figure*}
\includegraphics{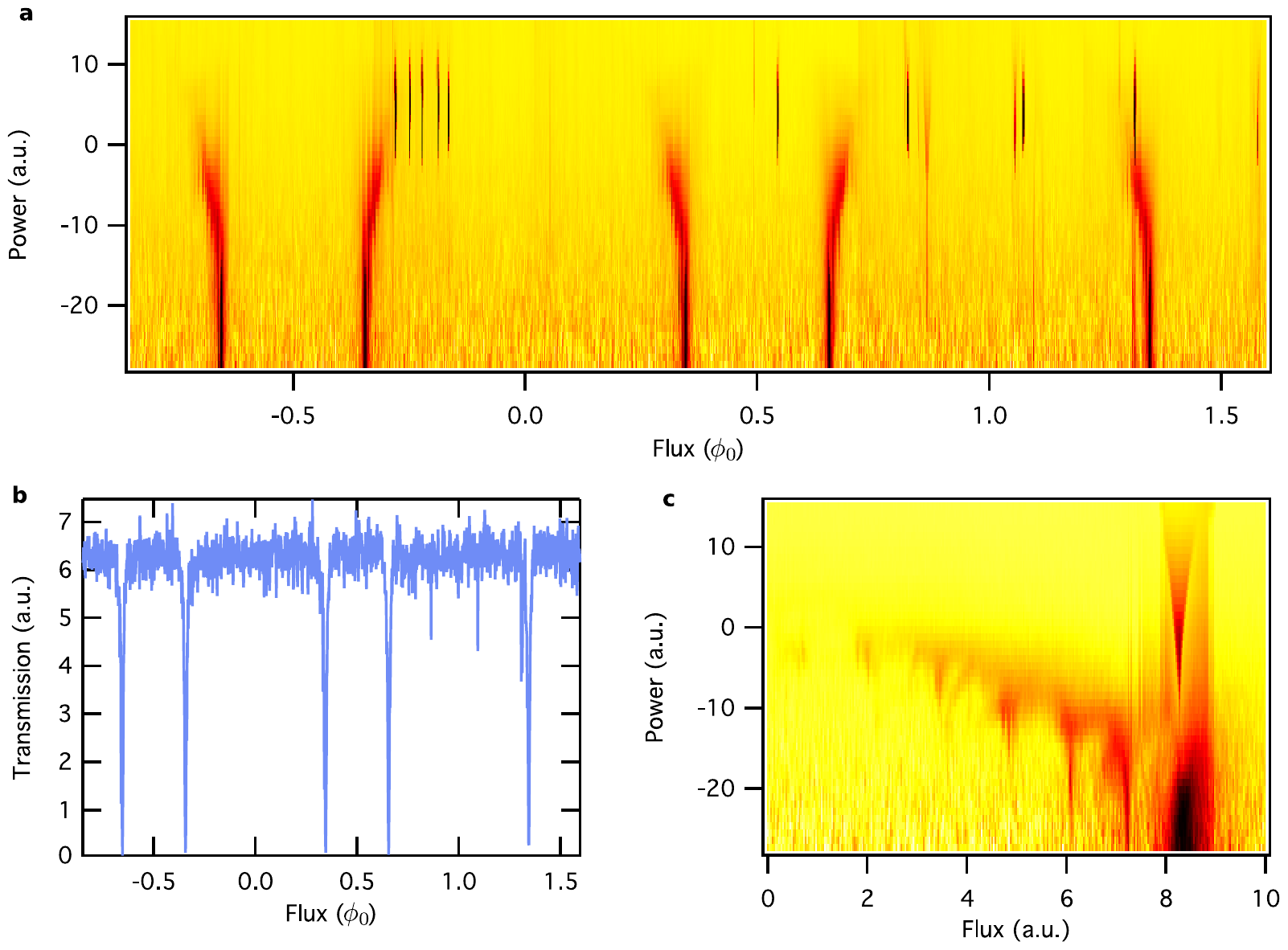}

\caption{\label{fig:finding_resonance}Transmission at the resonator frequency
versus input power and magnetic flux. \textsf{\textbf{a,}} The transmission
at $\nu_{r}$ is plotted versus input microwave power on a log scale
and magnetic flux threading the qubit loop. The flux axis has been
scaled by the observed flux period $\phi_{0}$. The power axis is
uncalibrated, but the cross-over near -15 on resonance occurs as the
photon occupation of the resonator increases from 0.1 to 10. The scan
was taken at the same $x$ position as Fig. \ref{fig:resonant_transmission_vs_y}
with the qubit close to the center pin and $g=29\,\text{MHz}$. \textsf{\textbf{b,}}
The average of the transmission for powers less than $ $-19 in panel\textsf{\textbf{
a}} is plotted versus flux. \textsf{\textbf{c,}} Another scan of transmission
versus power and flux like that in panel \textsf{\textbf{a}} is plotted
for a flux region close to resonance. The flux axis is uncalibrated.
The scan was taken at a position with $g=21\,\text{MHz}$, close to
the position used for the scan shown in panel \textsf{\textbf{a}}
.}
\end{figure*}

When $|\nu_{r}-\nu_{q}|\gg g$, one of the two mode frequencies $\nu_{\pm}$
given in equation (\ref{eq:mode_frequencies}) differs from $\nu_{r}$
by $\sim g^{2}/(\nu_{r}-\nu_{q})$ and is associated with a large
peak in transmission. For most qubit frequencies this frequency shift
is small compared to the resonator linewidth $\kappa=13\,\text{MHz}$
and so transmission at $\nu_{r}$ is high. However, when $ $$\nu_{q}\sim\nu_{r}$,
the mode frequencies $\nu_{\pm}$ are shifted from $\nu_{r}$ by $g>\kappa$
and transmission at $\nu_{r}$ is low. Panel \textsf{\textbf{a}} of
Fig. \ref{fig:finding_resonance} illustrates this behavior by showing
showing transmission at $\nu_{r}$ versus flux and input power. Regular
dips in transmission occur at low power where the qubit passes through
resonance. Panel \textsf{\textbf{b}} of Fig. \ref{fig:finding_resonance}
plots just the low power transmission versus flux and shows that resonance
can be easily identified by monitoring transmission at just one frequency
value. In practice, a scan like that shown in panel \textsf{\textbf{b}}
was taken at each position to identify the resonant flux range, and
then a scan like that shown in panel \textsf{\textbf{b}} of \ref{fig:transmission_fit}
was taken over this flux range to obtain transmission spectra to fit
for $g$. 

Additional features are present in the crossover from the low power
region to the high power region of the transmission. Panel \textsf{\textbf{c}}
of \ref{fig:transmission_fit} shows a finer scan of transmission
versus power and flux at a qubit position close to that of the scan
in panel \textsf{\textbf{a}}. These features are likely related to
higher level qubit transitions coming into resonance with the resonator,
though additional analysis is needed.

In the main text, results have been presented for the coupling of
the qubit to the resonator as a function of lateral position ($x$
and $y$). Measurements of the coupling's dependence on the qubit's
vertical displacement $z$ from the resonator were not possible due
to misalignment between the resonator and qubit chips. Evidence of
this misalignment is visible in panel \textsf{\textbf{a}} of Fig.
\ref{fig:finding_resonance} which plots the resonator frequency versus
the positioner's $z$ reading denoted by $z_{p}$. The origin of $z_{p}$
was chosen to be the point at which the positioner could no longer
advance due to contact with the resonator chip. Above $z_{p}=40\,\text{\textmu m}$,
the resonator frequency is shifted to higher values as the qubit chip
is brought closer as expected due to the modification of the resonator's
effective dielectric constant by the qubit's presence. Below $40\,\text{\textmu m}$,
the resonator frequency's dependence on $z_{p}$ weakens and disappears
even as the positioner continues to move. We interpret this behavior
as the qubit chip coming into first partial contact with the resonator
and then nearly full contact as compliance in the sample holder allow
the two chips to align. We attribute the relatively small magnitude
of the discrepancy of the qubit height obtained by the fit shown in
Fig. \ref{fig:g_fits} from the height of the photoresist pads to
this compliance in the sample holder. 

\begin{figure*}
\includegraphics{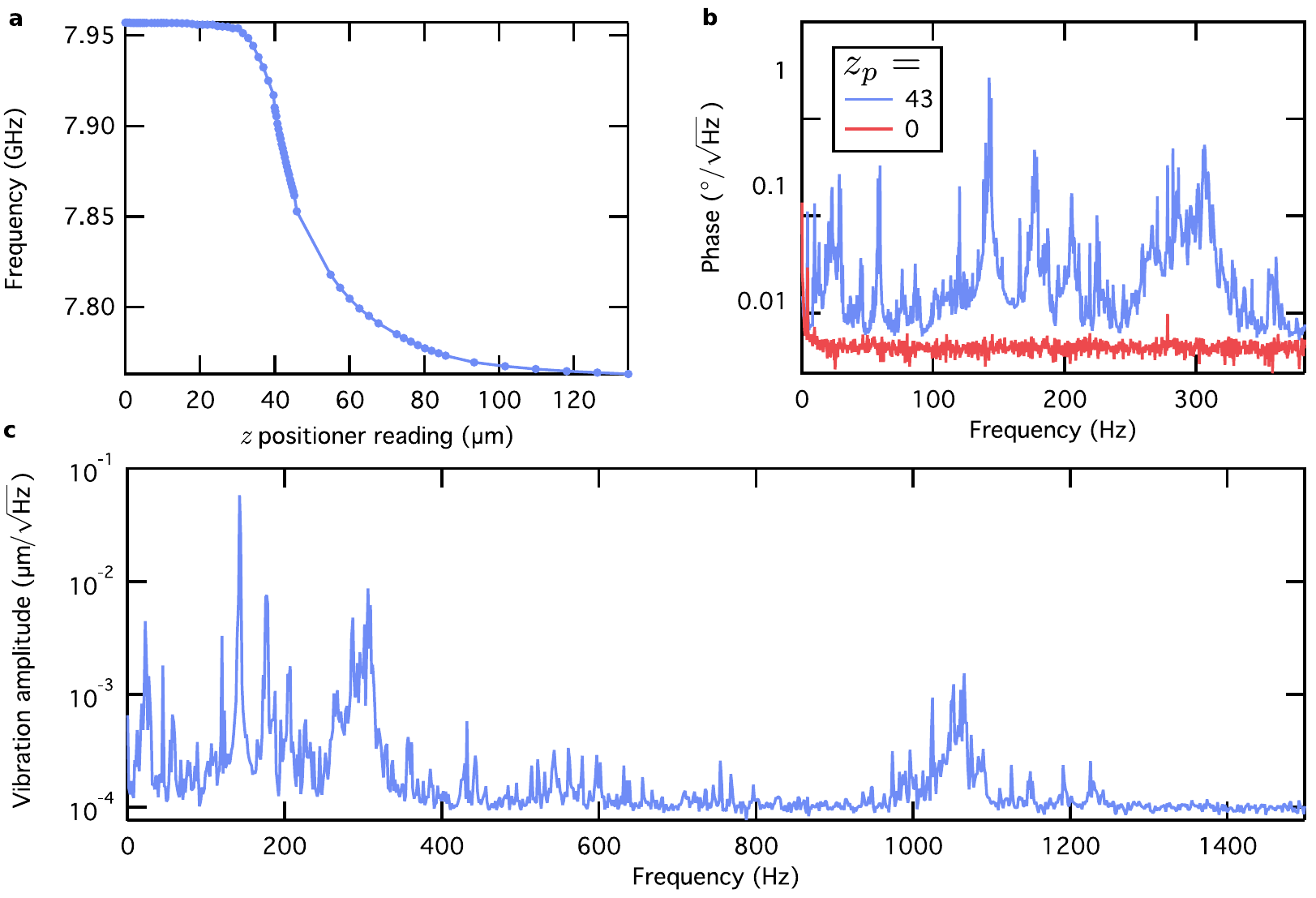}

\caption{\label{fig:vibrations}The effects of retracting the qubit from the
resonator chip. \textsf{\textbf{a,}} The resonator frequency is plotted
versus the reading $z_{p}$ of the $z$ positioner. At $z_{p}=0$,
the qubit chip is in hard contact with the resonator chip and can
not be advanced further. \textsf{\textbf{b,}} The noise spectrum of
the transmitted phase at the resonator frequency $\nu_{r}$ is plotted
for the positions $z_{p}=0$ and $43\,\text{\textmu m}$. \textsf{\textbf{c,}}
The noise spectrum of the qubit-resonator displacement is plotted
for $z_{p}=43\,\text{\textmu m}$ using the data from panel \textsf{\textbf{b}}.
The slope of the curve shown in panel \textsf{\textbf{a}} was used
to convert the phase data into displacement.}
\end{figure*}

We note that the shift of the resonator frequency due to the qubit
chip shown in panel \textsf{\textbf{a}} of Fig. \ref{fig:vibrations}
demonstrates the possibility of using a scannable chip to produce
a defect in a CQED lattice by shifting the frequency of one of the
resonators within the lattice. For scanning experiments where no defect
is desired, the shift shown in panel \textsf{\textbf{a}} could be
made unimportant by scanning with a chip larger than the entire CQED
lattice. In this case, the resonator at each lattice site would receive
the same shift.

The resonator frequency's dependence on the qubit chip height shown
in panel \textsf{\textbf{a}} of Fig. \ref{fig:vibrations} allowed
the resonator to be used to measure the qubit height. Panel \textsf{\textbf{b}}
of Fig. \ref{fig:vibrations} plots noise spectra of the transmitted
phase at the resonator frequency when the qubit chip is at $z_{p}=0$
and $43\,\text{\textmu m}$. We interpret the phase fluctuations present
when the qubit chip is hovering above the resonator and not present
when the qubit chip is in hard contact as being due to motion of the
qubit chip relative to the resonator chip. Using the resonator's phase
versus frequency curve to convert the phase noise into an effective
resonator frequency noise and then the curve shown in panel \textsf{\textbf{a}}
of Fig. \ref{fig:vibrations} to convert frequency into position,
we obtain the position noise spectrum shown in panel \textsf{\textbf{c}}
of Fig. \ref{fig:vibrations}. Repeating this procedure at another
$z_{p}$ position and using the amplitude noise instead of the phase
noise resulted in vibration spectra with similar features at similar
magnitudes, confirming the interpretation of the features as being
due to vibration of the qubit chip. The spectrum shown in panel \textsf{\textbf{c}}
of Fig. \ref{fig:vibrations} is typical for the mechanical response
of cryogenic positioners such as those used in this experiment. The
motion of the refrigerator base plate inferred from the spectrum shown
in panel \textsf{\textbf{c}} of Fig. \ref{fig:vibrations} agreed
with measurements made with an accelerometer of another refrigerator
that was the same model as that used for the measurements presented
here. Because all measurements of the qubit were made with the resonator
and qubit chips in hard contact, no vibration isolation elements were
included in the sample holder. In order to measure the dependence
of the qubit-resonator coupling on height, such vibration isolation
would need to be considered in addition to the alignment of the qubit
and resonator chips.

\def\maindoc{Scan_probe_g_vs_xy}
\newread\auxfile
\openin\auxfile = \maindoc .aux

\newcount\bibcounter
\bibcounter=0
\long\def\bibcitecheck#1#2\bibcitecheckstop{%
\ifx#1\bibcite
\advance \bibcounter by 1
\fi
}

\newif\ifnoteof
\loop
\read\auxfile to \holder

\ifeof\auxfile
\noteoffalse
\else
\expandafter\bibcitecheck\holder...\bibcitecheckstop
\noteoftrue
\fi

\ifnoteof
\repeat

\closein\auxfile

\makeatletter
\apptocmd{\thebibliography}{\global\c@NAT@ctr \bibcounter\relax}{}{}
\makeatother


\begin{thebibliography}{10}
\expandafter\ifx\csname url\endcsname\relax
  \def\url#1{\texttt{#1}}\fi
\expandafter\ifx\csname urlprefix\endcsname\relax\def\urlprefix{URL }\fi
\providecommand{\bibinfo}[2]{#2}
\providecommand{\eprint}[2][]{\url{#2}}

\bibitem{bloch2012quantum}
\bibinfo{author}{Bloch, I.}, \bibinfo{author}{Dalibard, J.} \&
  \bibinfo{author}{Nascimb{\`e}ne, S.}
\newblock \bibinfo{title}{Quantum simulations with ultracold quantum gases}.
\href {\doibase 10.1038/nphys2259} {
\newblock \emph{\bibinfo{journal}{Nature Phys.}} \textbf{\bibinfo{volume}{8}},
  \bibinfo{pages}{267--276} (\bibinfo{year}{2012}).
}

\bibitem{blatt2012quantum}
\bibinfo{author}{Blatt, R.} \& \bibinfo{author}{Roos, C.~F.}
\newblock \bibinfo{title}{Quantum simulations with trapped ions}.
\href {\doibase 10.1038/nphys2252} {
\newblock \emph{\bibinfo{journal}{Nature Phys.}} \textbf{\bibinfo{volume}{8}},
  \bibinfo{pages}{277--284} (\bibinfo{year}{2012}).
}

\bibitem{houck2012onchipquantum}
\bibinfo{author}{Houck, A.~A.}, \bibinfo{author}{T{\"u}reci, H.~E.} \&
  \bibinfo{author}{Koch, J.}
\newblock \bibinfo{title}{On-chip quantum simulation with superconducting
  circuits}.
\href {\doibase 10.1038/nphys2251} {
\newblock \emph{\bibinfo{journal}{Nature Phys.}} \textbf{\bibinfo{volume}{8}},
  \bibinfo{pages}{292--299} (\bibinfo{year}{2012}).
}

\bibitem{chiorescu2004coherent}
\bibinfo{author}{Chiorescu, I.} \emph{et~al.}
\newblock \bibinfo{title}{Coherent dynamics of a flux qubit coupled to a
  harmonic oscillator}.
\href {\doibase 10.1038/nature02831} {
\newblock \emph{\bibinfo{journal}{Nature}} \textbf{\bibinfo{volume}{431}},
  \bibinfo{pages}{159--162} (\bibinfo{year}{2004}).
}

\bibitem{wallraff2004strongcoupling}
\bibinfo{author}{Wallraff, A.} \emph{et~al.}
\newblock \bibinfo{title}{Strong coupling of a single photon to a
  superconducting qubit using circuit quantum electrodynamics}.
\href {\doibase 10.1038/nature02851} {
\newblock \emph{\bibinfo{journal}{Nature}} \textbf{\bibinfo{volume}{431}},
  \bibinfo{pages}{162--167} (\bibinfo{year}{2004}).
}

\bibitem{schoelkopf2008wiringup}
\bibinfo{author}{Schoelkopf, R.~J.} \& \bibinfo{author}{Girvin, S.~M.}
\newblock \bibinfo{title}{Wiring up quantum systems}.
\href {\doibase 10.1038/451664a} {
\newblock \emph{\bibinfo{journal}{Nature}} \textbf{\bibinfo{volume}{451}},
  \bibinfo{pages}{664--669} (\bibinfo{year}{2008}).
}

\bibitem{clarke2008superconducting}
\bibinfo{author}{Clarke, J.} \& \bibinfo{author}{Wilhelm, F.~K.}
\newblock \bibinfo{title}{Superconducting quantum bits}.
\href {\doibase 10.1038/nature07128} {
\newblock \emph{\bibinfo{journal}{Nature}} \textbf{\bibinfo{volume}{453}},
  \bibinfo{pages}{1031--1042} (\bibinfo{year}{2008}).
}

\bibitem{martinis2009superconducting}
\bibinfo{author}{Martinis, J.~M.}
\newblock \bibinfo{title}{Superconducting phase qubits}.
\href {\doibase 10.1007/s11128-009-0105-1} {
\newblock \emph{\bibinfo{journal}{Quantum Inf. Process.}}
  \textbf{\bibinfo{volume}{8}}, \bibinfo{pages}{81--103}
  (\bibinfo{year}{2009}).
}

\bibitem{blais2004cavityquantum}
\bibinfo{author}{Blais, A.}, \bibinfo{author}{Huang, R.-S.},
  \bibinfo{author}{Wallraff, A.}, \bibinfo{author}{Girvin, S.~M.} \&
  \bibinfo{author}{Schoelkopf, R.~J.}
\newblock \bibinfo{title}{Cavity quantum electrodynamics for superconducting
  electrical circuits: An architecture for quantum computation}.
\href {\doibase 10.1103/PhysRevA.69.062320} {
\newblock \emph{\bibinfo{journal}{Phys. Rev. A}}\,\,\,\, \textbf{\bibinfo{volume}{69}},
  \bibinfo{pages}{062320} (\bibinfo{year}{2004}).
}

\bibitem{greentree2006quantum}
\bibinfo{author}{Greentree, A.~D.}, \bibinfo{author}{Tahan, C.},
  \bibinfo{author}{Cole, J.~H.} \& \bibinfo{author}{Hollenberg, L. C.~L.}
\newblock \bibinfo{title}{Quantum phase transitions of light}.
\href {\doibase 10.1038/nphys466} {
\newblock \emph{\bibinfo{journal}{Nature Phys.}} \textbf{\bibinfo{volume}{2}},
  \bibinfo{pages}{856--861} (\bibinfo{year}{2006}).
}

\bibitem{hartmann2006strongly}
\bibinfo{author}{Hartmann, M.~J.}, \bibinfo{author}{Brand{\~a}o, F. G. S.~L.}
  \& \bibinfo{author}{Plenio, M.~B.}
\newblock \bibinfo{title}{Strongly interacting polaritons in coupled arrays of
  cavities}.
\href {\doibase 10.1038/nphys462} {
\newblock \emph{\bibinfo{journal}{Nature Phys.}} \textbf{\bibinfo{volume}{2}},
  \bibinfo{pages}{849--855} (\bibinfo{year}{2006}).
}

\bibitem{angelakis2007photonblockadeinduced}
\bibinfo{author}{Angelakis, D.~G.}, \bibinfo{author}{Santos, M.~F.} \&
  \bibinfo{author}{Bose, S.}
\newblock \bibinfo{title}{Photon-blockade-induced Mott transitions and {XY}
  spin models in coupled cavity arrays}.
\href {\doibase 10.1103/PhysRevA.76.031805} {
\newblock \emph{\bibinfo{journal}{Phys. Rev. A}}\,\,\, \textbf{\bibinfo{volume}{76}},
  \bibinfo{pages}{031805} (\bibinfo{year}{2007}).
}

\bibitem{greiner2002quantum}
\bibinfo{author}{Greiner, M.}, \bibinfo{author}{Mandel, O.},
  \bibinfo{author}{Esslinger, T.}, \bibinfo{author}{H{\"a}nsch, T.~W.} \&
  \bibinfo{author}{Bloch, I.}
\newblock \bibinfo{title}{Quantum phase transition from a superfluid to a Mott
  insulator in a gas of ultracold atoms}.
\href {\doibase 10.1038/415039a} {
\newblock \emph{\bibinfo{journal}{Nature}} \textbf{\bibinfo{volume}{415}},
  \bibinfo{pages}{39--44} (\bibinfo{year}{2002}).
}

\bibitem{hayward2012fractional}
\bibinfo{author}{Hayward, A. L.~C.}, \bibinfo{author}{Martin, A.~M.} \&
  \bibinfo{author}{Greentree, A.~D.}
\newblock \bibinfo{title}{Fractional quantum Hall physics in
  Jaynes-Cummings-Hubbard lattices}.
\href {\doibase 10.1103/PhysRevLett.108.223602} {
\newblock \emph{\bibinfo{journal}{Phys. Rev. Lett.}} 
  \textbf{\bibinfo{volume}{108}}, \bibinfo{pages}{223602}
  \,\,\,\,\,\,\,(\bibinfo{year}{2012}).
}

\bibitem{petrescu2012anomalous}
\bibinfo{author}{Petrescu, A.}, \bibinfo{author}{Houck, A.~A.} \&
  \bibinfo{author}{Le~Hur, K.}
\newblock \bibinfo{title}{Anomalous Hall effects of light and chiral edge modes
  on the kagom{\'e} lattice}.
\href {\doibase 10.1103/PhysRevA.86.053804} {
\newblock \emph{\bibinfo{journal}{Phys. Rev. A}} \textbf{\bibinfo{volume}{86}},
  \bibinfo{pages}{053804} (\bibinfo{year}{2012}).
}

\bibitem{bardyn2012majoranalike}
\bibinfo{author}{Bardyn, C.-E.} \& \bibinfo{author}{{\.I}mamo{\v g}lu, A.}
\newblock \bibinfo{title}{Majorana-like modes of light in a one-dimensional
  array of nonlinear cavities}.\\
\href {\doibase 10.1103/PhysRevLett.109.253606} {
\newblock \emph{\bibinfo{journal}{Phys. Rev. Lett.}}
  \textbf{\bibinfo{volume}{109}}, \bibinfo{pages}{253606}
  (\bibinfo{year}{2012}).
}

\bibitem{underwood2012lowdisorder}
\bibinfo{author}{Underwood, D.~L.}, \bibinfo{author}{Shanks, W.~E.},
  \bibinfo{author}{Koch, J.} \& \bibinfo{author}{Houck, A.~A.}
\newblock \bibinfo{title}{Low-disorder microwave cavity lattices for quantum
  simulation with photons}.
\href {\doibase 10.1103/PhysRevA.86.023837} {
\newblock \emph{\bibinfo{journal}{Phys. Rev. A}} \textbf{\bibinfo{volume}{86}},
  \bibinfo{pages}{023837} (\bibinfo{year}{2012}).
}

\bibitem{eriksson1996cryogenic}
\bibinfo{author}{Eriksson, M.~A.} \emph{et~al.}
\newblock \bibinfo{title}{Cryogenic scanning probe characterization of
  semiconductor nanostructures}.
\href {\doibase doi:10.1063/1.117801} {
\newblock \emph{\bibinfo{journal}{Appl. Phys. Lett.}}
  \textbf{\bibinfo{volume}{69}}, \bibinfo{pages}{671--673}
  (\bibinfo{year}{1996}).
}

\bibitem{topinka2000imaging}
\bibinfo{author}{Topinka, M.~A.} \emph{et~al.}
\newblock \bibinfo{title}{Imaging coherent electron flow from a quantum point
  contact}.
\href {\doibase 10.1126/science.289.5488.2323} {
\newblock \emph{\bibinfo{journal}{Science}} \textbf{\bibinfo{volume}{289}},
  \bibinfo{pages}{2323--2326} (\bibinfo{year}{2000}).
}

\bibitem{johnson2010quantum}
\bibinfo{author}{Johnson, B.~R.} \emph{et~al.}
\newblock \bibinfo{title}{Quantum non-demolition detection of single microwave
  photons in a circuit}.
\href {\doibase 10.1038/nphys1710} {
\newblock \emph{\bibinfo{journal}{Nature Phys.}} \textbf{\bibinfo{volume}{6}},
  \bibinfo{pages}{663--667} (\bibinfo{year}{2010}).
}

\bibitem{rondin2012nanoscale}
\bibinfo{author}{Rondin, L.} \emph{et~al.}
\newblock \bibinfo{title}{Nanoscale magnetic field mapping with a single spin
  scanning probe magnetometer}.
\href {\doibase doi:10.1063/1.3703128} {
\newblock \emph{\bibinfo{journal}{Appl. Phys. Lett.}}
  \textbf{\bibinfo{volume}{100}}, \bibinfo{pages}{153118}
  (\bibinfo{year}{2012}).
}

\bibitem{maletinsky2012arobust}
\bibinfo{author}{Maletinsky, P.} \emph{et~al.}
\newblock \bibinfo{title}{A robust scanning diamond sensor for nanoscale
  imaging with single nitrogen-vacancy centres}.
\href {\doibase 10.1038/nnano.2012.50} {
\newblock \emph{\bibinfo{journal}{Nat. Nanotechnol.}}
  \textbf{\bibinfo{volume}{7}}, \bibinfo{pages}{320--324}
  (\bibinfo{year}{2012}).
}

\bibitem{englund2010deterministic}
\bibinfo{author}{Englund, D.} \emph{et~al.}
\newblock \bibinfo{title}{Deterministic coupling of a single nitrogen vacancy
  center to a photonic crystal cavity}.
\href {\doibase 10.1021/nl101662v} {
\newblock \emph{\bibinfo{journal}{Nano Lett.}} \textbf{\bibinfo{volume}{10}},
  \bibinfo{pages}{3922--3926} (\bibinfo{year}{2010}).
}

\bibitem{koch2007chargeinsensitive}
\bibinfo{author}{Koch, J.} \emph{et~al.}
\newblock \bibinfo{title}{Charge-insensitive qubit design derived from the
  Cooper pair box}.
\href {\doibase 10.1103/PhysRevA.76.042319} {
\newblock \emph{\bibinfo{journal}{Phys. Rev. A}} \textbf{\bibinfo{volume}{76}},
  \bibinfo{pages}{042319} (\bibinfo{year}{2007}).
}

\bibitem{huber2008gradiometric}
\bibinfo{author}{Huber, M.~E.} \emph{et~al.}
\newblock \bibinfo{title}{Gradiometric micro-{SQUID} susceptometer for scanning
  measurements of mesoscopic samples}.
\href {\doibase doi:10.1063/1.2932341} {
\newblock \emph{\bibinfo{journal}{Rev. Sci. Instrum.}}
  \textbf{\bibinfo{volume}{79}}, \bibinfo{pages}{053704}
  (\bibinfo{year}{2008}).
}

\bibitem{bialczak2011fasttunable}
\bibinfo{author}{Bialczak, R.~C.} \emph{et~al.}
\newblock \bibinfo{title}{Fast tunable coupler for superconducting qubits}.
\href {\doibase 10.1103/PhysRevLett.106.060501} {
\newblock \emph{\bibinfo{journal}{Phys. Rev. Lett.}}
  \textbf{\bibinfo{volume}{106}}, \bibinfo{pages}{060501}
  \,\,\,\,\,\,\,\,\,\,\,\,\,\,\,(\bibinfo{year}{2011}).
}

\bibitem{srinivasan2011tunable}
\bibinfo{author}{Srinivasan, S.~J.}, \bibinfo{author}{Hoffman, A.~J.},
  \bibinfo{author}{Gambetta, J.~M.} \& \bibinfo{author}{Houck, A.~A.}
\newblock \bibinfo{title}{Tunable coupling in circuit quantum electrodynamics
  using a superconducting charge qubit with a V-shaped energy level diagram}.
\href {\doibase 10.1103/PhysRevLett.106.083601} {
\newblock \emph{\bibinfo{journal}{Phys. Rev. Lett.}}
  \textbf{\bibinfo{volume}{106}}, \bibinfo{pages}{083601}
  (\bibinfo{year}{2011}).
}

\bibitem{dicarlo2009demonstration}
\bibinfo{author}{{DiCarlo}, L.} \emph{et~al.}
\newblock \bibinfo{title}{Demonstration of two-qubit algorithms with a
  superconducting quantum processor}.
\href {\doibase 10.1038/nature08121} {
\newblock \emph{\bibinfo{journal}{Nature}} \textbf{\bibinfo{volume}{460}},
  \bibinfo{pages}{240--244} (\bibinfo{year}{2009}).
}

\end{thebibliography}

\begin{thebibliography}{31}
\expandafter\ifx\csname url\endcsname\relax
  \def\url#1{\texttt{#1}}\fi
\expandafter\ifx\csname urlprefix\endcsname\relax\def\urlprefix{URL }\fi
\providecommand{\bibinfo}[2]{#2}
\providecommand{\eprint}[2][]{\url{#2}}

\bibitem{schreier2008suppressing}
\bibinfo{author}{Schreier, J.~A.} \emph{et~al.}
\newblock \bibinfo{title}{Suppressing charge noise decoherence in
  superconducting charge qubits}.
\href {\doibase 10.1103/PhysRevB.77.180502} {
\newblock \emph{\bibinfo{journal}{Phys. Rev. B}} \textbf{\bibinfo{volume}{77}},
  \bibinfo{pages}{180502} (\bibinfo{year}{2008}).
}

\end{thebibliography}
\end{document}